\documentclass[letterpaper,journal]{IEEEtran}
\usepackage[dvips]{graphicx}
\usepackage{amsmath,amsfonts,amssymb}
\usepackage{verbatim}

\usepackage{theorem}

\newtheorem{theorem}{Theorem}

\def\squareforqed{\hbox{\rlap{$\sqcap$}$\sqcup$}}
\def\qed{\ifmmode\squareforqed\else{\unskip\nobreak\hfil
\penalty50\hskip1em\null\nobreak\hfil\squareforqed
\parfillskip=0pt\finalhyphendemerits=0\endgraf}\fi}
\def\endenv{\ifmmode\;\else{\unskip\nobreak\hfil
\penalty50\hskip1em\null\nobreak\hfil\;
\parfillskip=0pt\finalhyphendemerits=0\endgraf}\fi}

\newcommand{\ket}[1]{\mathop{\left|#1\right>}\nolimits}       

\newcommand{\kb}[2]{| #1\rangle\!\langle #2 |}

\newcommand{\Li}[2]{{\mathrm{Li}}\left(#1,#2\right)}

\newcommand{\openone}{{\mathbb I}}

\newcommand{\nc}{\newcommand}
\nc{\rar}{\rightarrow}
\nc{\lrar}{\longrightarrow}
\def\a{\alpha}
\def\b{\beta}

\def\d{\delta}
\def\e{\epsilon}

\def\r{\rho}
\def\s{\sigma}

\def\ph{\varphi}

\def\ps{\psi}
\def\o{\omega}

\def\Ph{\Phi}

\nc{\cB}{{\cal B}}
\nc{\cD}{{\cal D}}
\nc{\cE}{{\cal E}}
\nc{\cM}{{\cal M}}
\nc{\cN}{{\cal N}}
\nc{\cR}{{\cal R}}

\nc{\ox}{\otimes}
\nc{\dg}{\dagger}
\nc{\proj}[1]{| #1\rangle\!\langle #1 |}
\nc{\id}{\operatorname{id}}
\nc{\smfrac}[2]{\mbox{$\frac{#1}{#2}$}}

\nc{\vac}{\mathsf{vac}}

\begin{document}

\title{Private information via the Unruh effect}

\author{
    Kamil Br\'adler$^\dagger$
    Patrick Hayden$^\dagger$
    Prakash Panangaden$^\dagger$
    \thanks{$\dagger$
        School of Computer Science, McGill University, Montr\'{e}al, Canada}
}

\date{May 2, 2009}


\maketitle

\begin{abstract}
In a relativistic theory of quantum information, the possible
presence of horizons is a complicating feature placing restrictions
on the transmission and retrieval of information. We consider two
inertial participants communicating via a noiseless qubit channel in
the presence of a uniformly accelerated eavesdropper. Owing to the
Unruh effect, the eavesdropper's view of any encoded information is
noisy, a feature the two inertial participants can exploit to
achieve perfectly secure quantum communication. We show that the
associated private quantum capacity is equal to the
entanglement-assisted quantum capacity for the channel to the
eavesdropper's environment, which we evaluate for all accelerations.
\end{abstract}


\PARstart{Q}{\MakeLowercase {uantum}} information theory for the
most part assumes that the senders, receivers and eavesdroppers
involved in a protocol share an inertial frame. For many of the
applications envisioned in the field this is a good approximation
and sometimes, as in the case of quantum key distribution, even a
rigorously justifiable simplification. To the extent that quantum
information theory attempts to identify fundamental rules governing
information processing imposed by the laws of physics , however,
neglecting relativity is ultimately unacceptable. Luckily, much of
the formalism of quantum information remains valid in relativistic
settings and the effect of changing reference frames can usually be
modeled as the introduction of noise. Thus, there has been a
significant amount of work done to calculate how entanglement
degrades under a boost or
acceleration~\cite{CW03,Peres04,gingadami,Ahnetal03,cabremb05,fuentes-schuller:120404}
and how basic quantum information theoretic protocols like
teleportation, which were designed for inertial participants, break
down under acceleration~\cite{Alsing03}.

The natural next step is to design communications protocols
specifically adapted to relativistic situations and, possibly, take
advantage of uniquely relativistic features to accomplish otherwise
impossible tasks. Kent has demonstrated, for example, that secure
bit commitment is possible using a protocol exploiting relativistic
causality constraints even though it is known to be impossible
otherwise~\cite{Kent99}. In this article, we consider a scenario in
which two inertial participants communicate via a noiseless,
bosonic, dual-rail qubit channel in the presence of a uniformly
accelerated eavesdropper. In this context, the eavesdropper's Unruh
noise becomes a resource which the inertial participants can
potentially exploit to encrypt their communications.

The \emph{private quantum capacity} is the optimal rate at which a
sender (Alice) can send \emph{qubits} to a receiver (Bob) while
keeping them private from an eavesdropper (Eve). It had not
previously been studied because in most settings it is redundant to
require privacy in quantum communication: if the eavesdropper is
modeled as being part of the environment of the communications
channel then quantum communication is automatically private. This
was in fact the insight behind Devetak's proof of the quantum
capacity theorem~\cite{Devetak05}. On the other hand, if Eve is
assumed to have unrestricted access to the states while they are in
transit from Alice to Bob, then unconditional privacy is impossible
without secret keys in a nonrelativistic model because Eve and Bob
are effectively interchangeable. This symmetry is broken, however,
if Eve is assumed to be accelerating. The private quantum capacity
problem therefore provides an example of a question which is poorly
motivated in non-relativistic settings but very natural when
relativity is taken into account.
%
%
Because of structural features of the Unruh effect, this private
quantum capacity is exactly zero if Alice is restricted to isometric
encodings. However, for general encodings it is given by the same
formula as the entanglement-assisted quantum capacity~\cite{BSST02}
of the channel to the eavesdropper's environment, despite the
absence of any operational connection between the two
problems.
%
%
Of course, it is also possible to define a private classical
capacity for this setting, which we study for the purposes of
comparison with its quantum version.

\section{Unruh channels} \label{sec:channel}

The Unruh effect, whereby an observer uniformly accelerated in a
vacuum experiences a thermal bath~\cite{Unruh76,UnruhReview}, can be
understood as a consequence of the fact that an accelerated observer
has a different Fock representation of the quantum field than does
an inertial observer. In particular, the vacuum state as defined by
an inertial observer will be a thermal state in the Fock space
defined by a uniformly accelerating observer. The transformation
between these Fock spaces is conveniently represented by a
transformation of creation and annihilation operators called a
Bogoliubov transformation~\cite{WaldBook}.

Consider a state $\ket{\psi}$ of a quantum field.  The inertial
observers may see this as a many particle state: $\Pi_i
a_i^{\dg}\ket{\vac}$.  The Bogoliubov transformation changes each
$a_i$ to some combination of the creation and annihilation operators
of the non-inertial observer's Fock decomposition. In our case, the
Bogoliubov transformation, which relates the Fock decompositions in
the Minkowski and Rindler spacetimes corresponding respectively to
the inertial and accelerating observers, has a very special form
because of the spacetime symmetries.  The only mixing of modes is
between the modes with the same momentum in the two Rindler wedges.
The \emph{Unruh channel} $\cN$ consists of this change composed with
tracing over the modes that are inaccessible to the accelerating
observer because they are beyond her horizon.

\begin{figure}[t]
\begin{center}
\includegraphics[width=6.5cm]{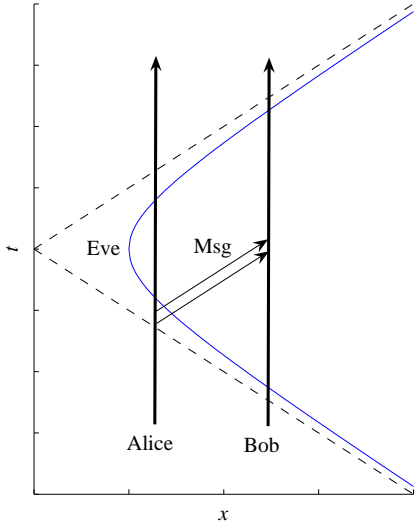}
\end{center}
\caption{Spacetime diagram of the communication scenario. Alice and
Bob are inertial observers, assumed without loss of generality to be
at rest. Meanwhile, Eve is uniformly accelerated resulting in future
and past horizons indicated by the dashed lines $x = \pm t$. Eve
intercepts a message sent from Alice to Bob but will be thwarted in
her attempts to determine the contents even though the only
restrictions considered on Eve's ability to perform quantum
measurements are those arising from the presence of the horizons.}
 \label{fig:rindler}
\end{figure}

We will assume that Alice encodes information for Bob by preparing
quantum states of a bosonic, dual-rail qubit as illustrated in
figure \ref{fig:rindler}. In other words, she has access to a
two-dimensional sector of her (and Bob's) Fock space, with basis
vectors given by a single excitation of a massless scalar field in
one of two different modes, which we label by their associated
annihilation operators $a$ and $b$.\footnote{Throughout the paper,
we work with plane wave modes which are not, strictly speaking,
physically realizable. Nonetheless, the superpositions involved in
defining a wave packet can be carried through our calculations using
approximate mode matching because the Bogoliubov transformation does
not mix modes. As a result, explicit calculations using wave packets
do not lead to substantial
differences for our purposes~\cite{Audretsch94,Bradler07}.}
$U_{ac}(r) = \exp[ r( a^\dg c^\dg - ac) ]$ is the unitary operator
transforming the sector of Alice's Fock space to the corresponding
sector of Eve's Fock space. (Because the Bogoliubov transformation
is diagonal, we can safely ignore all other modes~\cite[p.
106]{Mukhanov}.) In short, the channel is $U_{ac}$ followed by the
appropriate trace. The parameter $r$ is related to Eve's proper
acceleration $\tau$ and the mode frequency $\o$ by
$\tanh{r}=\exp{(-\pi\omega /\tau)}$~\cite{Alsing03,WaldBook}.

In our dual-rail case, an arbitrary pure input state $\ket{\psi} =
(\a b^\dg + \b a^\dg)\ket{\vac}$ is transformed to Eve's Fock space
according to $U_{abcd} = U_{ac} \ox U_{bd}$, which expands to
\begin{multline}\label{squeeze}
        U_{abcd}(r)
        =\smfrac{1}{\cosh^2{r}}e^{\tanh{r}(a^\dg c^\dg+b^\dg d^\dg)}\\
        \times e^{-\ln{\cosh{r}}(a^\dg a+b^\dg b+c^\dg c+d^\dg d)}e^{-\tanh{r}(ac+bd)}.
\end{multline}
For all states in the dual-rail basis, equation~(\ref{squeeze})
reduces to $U_{abcd}(r)=1/\cosh^3{r}\exp{[\tanh{r}(a^\dg c^\dg+b^\dg
d^\dg)]}$. This allows us to write the state in Eve's Fock space as
$\ket{\psi}=U_{abcd}(r)(\alpha b^\dg+\beta a^\dg)\ket{\vac}=(\alpha
b^\dg+\beta a^\dg)U_{abcd}(r)\ket{\vac}$. If we trace over degrees
of freedom beyond Eve's horizon ($cd$), then $\sigma =
\cN(\proj{\psi})= (1-z)^3\bigoplus_{k=0}^\infty z^k\,\sigma_k$ is
block diagonal with blocks $\sigma_k$ labeled by the total
excitation number $k$ ($z=\tanh^2{r}$):
\begin{multline}\label{Eve_state}
    \sigma_k = \sum_{n=0}^k\Bigl[
    |\alpha|^2(n+1)\kb{k-n,n+1}{k-n,n+1}\\
    +|\beta|^2(k-n+1)\kb{k-n+1,n}{k-n+1,n}\\
    +\alpha\bar\beta\sqrt{(n+1)(k-n+1)}\kb{k-n,n+1}{k-n+1,n}\\
    +h.c.\Bigl].
\end{multline}
Each block $\sigma_k$ can be expressed as a linear combination of
generators $J_x^{(k+2)}$, $J_y^{(k+2)}$ and $J_z^{(k+2)}$ of the
irreducible $(k+2)-$dimensional representation of $SU(2)$.
($\vec{J}^{(2)}$, for example, consists of the Pauli matrices scaled
by $1/2$.) If $\sigma = \cN(\rho)$ with $\rho = \openone/2+ \vec{n}
\cdot \vec{J}^{(2)}$ arbitrary, then
\begin{equation}
\sigma_k = \openone(k+1)/2+n_xJ_x^{(k+2)}+n_yJ_y^{(k+2)}+n_zJ_z^{(k+2)}.
\end{equation}
As a consequence, the channel $\cN$ to Eve is covariant in the sense
that $\cN(U \rho U^\dg) = R(U) \cN(\rho) R(U^\dg)$ where $R$ is the
infinite dimensional representation of $SU(2)$ given by the direct
sum over all its finite dimensional irreps. This makes it easy to
diagonalize $\s$: the eigenvalues of $\s_k$ are equally spaced with
spacing $S = \| \vec{n} \|_2$ and largest eigenvalue equal to
$(k+1)(S+1)/2$.

\section{Private quantum capacity} \label{sec:qpriv}


Capacities are defined by allowing arbitrarily many uses of a
channel and asking that the various data transmission or privacy
requirements hold to any desired level of approximation in the limit
of many uses.
The private quantum capacity is defined as the optimal rate at which
Alice can send qubits to Bob while simultaneously ensuring that
those qubits appear to be completely encrypted from Eve's point of
view.
%
There are several equivalent ways of formalizing this notion, but we
will take it to mean that Alice would like to transmit halves of
entangled pairs to Bob. Privacy in this context means that there
should be no correlation between the output of Eve's channel and the
halves of the entangled pairs kept in Alice's laboratory.
%
%
Since the private quantum capacity has not been studied elsewhere,
we begin by providing some formal definitions and studying the
general case.

Given are a quantum channel $\cN_1$ from Alice to Bob and another
$\cN_2$ from Alice to Eve. Let $\Phi_{2^k}$ represent the density
operator of $k$ maximally entangled pairs of qubits and $\tau_{2^k}$
the maximally mixed state on $k$ qubits. An $(n,k,\d,\e)$
\emph{private entanglement transmission code} from Alice to Bob
consists of an encoding channel $\cE$ taking $k$ qubits into the
input of $\cN_1^{\ox n}$ and a decoding channel $\cD$ taking the
output of Bob's channel $\cN_1^{\ox n}$ back to $k$ qubits
satisfying
\parbox[c]{3.2in}{
\smallskip
\noindent 1. \emph{Transmission:} \\
$
 \left\|
    (\id \ox \cD \circ \cN_1^{\ox n} \circ \cE)
    ( \Ph_{2^k} ) -
     \Ph_{2^k}
 \right\|_1 \leq \d.
 $
\smallskip

\noindent 2. \emph{Privacy:} \\
$
 \left\|
    (\id \ox \cN_2^{\ox n} \circ \cE)
    ( \Ph_{2^k} )   -
   \tau_{2^k} \, \ox (\cN_2^{\ox n} \circ \cE) (
   \tau_{2^k} )
 \right\|_1\leq \e.
 $
}
\smallskip

A rate $Q$ is an \emph{achievable} rate for private entanglement
transmission if for all $\d, \e > 0$ and sufficiently large $n$
there exist $(n,\lfloor nQ \rfloor, \d, \e)$ private entanglement
transmission codes. The private quantum capacity is the supremum of
the achievable rates.
For a density operator $\sigma^{AB}$, let $H(A)_\s$ be the von
Neumann entropy of $\s^A$. The mutual information $I(A;B)_\s$ is
$H(A)_\s + H(B)_\s - H(AB)_\s$.

\begin{theorem} \label{thm:qpriv}
The private quantum capacity $Q_p(\id,\cN)$ when the channel from
Alice to Bob is noiseless is given by the formula $\max
\smfrac{1}{2} I(A;E_c)_\r$, where the maximization is over all pure
states $\ket{\ph}^{AA'}$ and $\r = (\id \ox \cN_c)(\ph)$. $\cN_c$ is
the channel to Eve's environment $E_c$, that is, the complement of
$\cN$ with respect to its Stinespring dilation.
\end{theorem}
Despite the absence here of any entanglement assistance, the theorem
implies that $Q_p(\id_2,\cN)$ is exactly equal to the
entanglement-assisted quantum capacity of $\cN_c$, usually written
$Q_E(\cN_c)$~\cite{BSST02}.

\smallskip
To see that the advertised rate is achievable, write $V_\cE$ for the
Stinespring extension of $\cE$, with output on $A^{'n} F$. The
privacy condition applied to $\cN$ is equivalent to entanglement
transmission to $FE_c^n$ via Uhlmann's theorem~\cite{Nielsen00}. It
is therefore sufficient to find codes that simultaneously perform
entanglement transmission to Bob and to $FE_c^n$. The encodings
analyzed in \cite{Hayden08} do not depend on the channels, however,
just the dummy input $\ph$, so the same encodings can be employed
for both tasks. Choosing $|F|=2^{nf}$, the following sufficient
conditions for successful transmission can be extracted from
\cite{Hayden08,BBN08}:
\begin{equation}
Q < H(A)_\ph - f \quad\mbox{and}\quad Q < I(A\rangle E_c)_\r + f,
\end{equation}
where $I(A\rangle E_c)_\r$ is the coherent information $H(E_c)_\r -
H(AE_c)_\r$. These constraints have intuitive interpretations: the
first is the noiseless rate to Bob through $V_\cE$ reduced by the
rate at which qubits are lost to $F$, while the second is the
standard coherent information rate for $\cN_c$ augmented by a
noiseless channel to $F$. $Q$ is maximized subject to these
constraints when $H(A)_\ph - f = I(A\rangle E_c)_\r + f$. Using the
fact that $H(A)_\ph = H(A)_\r$ and purifying $\r$ to
$\ket{\r}^{AEE_c}$, this equation can be written as $f =
\smfrac{1}{2}I(A;E)_\r$. Therefore, the rate $Q$ is achievable
provided $Q < H(A)_\r - \smfrac{1}{2}I(A;E)_\r =
\smfrac{1}{2}I(A;E_c)_\r$.

To prove optimality, suppose we have an $(n, \lfloor nQ
\rfloor,\d,\e)$ private entanglement transmission code. Use $R$ to
denote the reference space for the maximally entangled state
$\Phi_{2^k}$ in the definition. Let $\ket{\s}^{RFE^nE_c^n}$ be the
purified final state after $\cN_2^{\ox n} \circ \cE$ has acted on
$\Ph_{2^k}$. The privacy condition and the Alicki-Fannes'
inequality~\cite{AF03} imply that there is a function $g(\d)$
satisfying $\lim_{\d \rar 0} g(\d) = 0$ such that
\begin{eqnarray}
2 \lfloor nQ \rfloor = 2 \log |R|
 &\leq& I(R;E_c^n F)_\s + n g(\d) \\
 &=& I(R;F)_\s + I(R;E_c^n|F)_\s \nonumber \\
 &\leq& I(R;E_c^n|F)_\s + n g(\d+\e) \nonumber \\
 &\leq& I(RF;E_c^n)_\s + n g(\d+\e). \nonumber
\end{eqnarray}
The first inequality is a consequence of the existence of the
decoding channel $\cD$ and the monotonicty of mutual information.
The second inequality holds because entanglement transmission to Bob
requires no leakage to $F$, leading to an upper bound on
$I(R;F)_\s$. The final inequality follows from the chain rule and
the nonnegativity of mutual information. Labeling $RF$ by $A$, we
can conclude that
\begin{equation}
Q_p(\id, \cN) \leq \lim_{n \rar \infty} \max \frac{1}{2n} I(A;E_c^n)_\r,
\end{equation}
where the maximization is pure states $\ket{\ph}^{A^nA'^n}$ and $\r
= (\id \ox \cN_c^{\ox n})(\ph)$. It is well-known, however, that
fixing $n=1$ does not affect the expression on the right hand side
of the inequality~\cite{BSST02},  finishing the proof of optimality.

\subsection{Unruh case}

Let us now focus on the the case where $\cN$ is the Unruh channel.
Inspection of figure \ref{fig:rindler} reveals that there is only a
finite amount of time during which Eve can intercept messages from
Alice to Bob. The limit $n \rar \infty$ of infinite length messages
considered in the definition of the private classical capacity
therefore does not formally apply, but codes nearly achieving the
capacity can be found for reasonably small $n$.

It is instructive to first consider encodings $\cE$ that are
isometric, a restriction that does not affect the regular
(non-private) quantum capacity. Private entanglement transmission
codes then simply become codes that transmit entanglement beyond
Eve's horizon to $E_c$.

Taking the partial trace over $(ab)$ instead of $(cd)$ of the pure
state $\ket{\psi}$ from Eve's Fock space yields the channel $\cN_c$
from Alice to $E_c$, the Hilbert space describing degrees of freedom
beyond Eve's horizon. The result is
\begin{equation}
\cN_c(\rho) = z \, \bar{\s} + (1 - z )\,  \o_0,
\end{equation}
where $\s = \cN(\rho)$ and $\o_0$ is a diagonal state independent of
$\rho$. Therefore, given the output $\sigma$ to her channel, Eve can
simulate the channel to $E_c$ up to complex conjugation. The
simulation is simple. With probability $z$ she does nothing
to $\sigma$ and with probability $1 - z$ she prepares $\o_0$
and uses it to replace $\s$.

Now suppose that it is possible to transmit entanglement (and
therefore quantum states) beyond Eve's horizon. Write $\cD(\tau) =
\sum_j D_j \tau D_j^\dg$ for the decoding channel on $E_c$. Since
$\bar{\cD}(\tau) = \sum_j \bar{D}_j \tau \bar{D}_j^\dg$ is also a
quantum channel, Eve can apply $\bar{\cD}$ to the output of her
simulation. Assuming high fidelity transmission of a quantum state
$\ket{\psi}$ beyond Eve's horizon, the output of $\bar{\cD}$ will be
a high fidelity transmission of $\ket{\bar{\psi}}$. That is
impossible because the map $\ket{\psi} \mapsto
\ket{\psi}\ket{\bar{\psi}}$, the result of applying both decodings
in parallel, is nonlinear. The only possible conclusion is that it
must be impossible to transmit entanglement beyond Eve's horizon.
It is therefore impossible to achieve private entanglement transmission using isometric codes.

\begin{figure}[t]
\begin{center}
\includegraphics[height=3.7cm,width=8.5cm,bb=25 280 575 515]{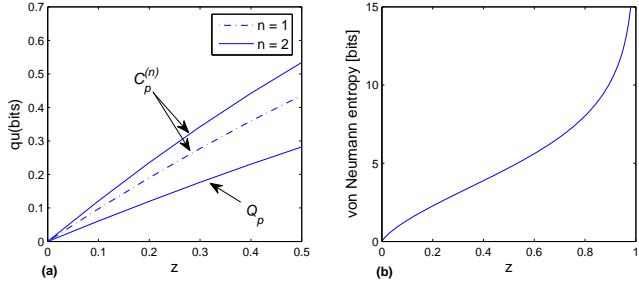}
\end{center}
\caption{{\bf (a)} The private quantum capacity $Q_p(\id_2,\cN)$ is
non-zero for all positive accelerations and strictly less than
$C_p^{(n)}$ for all $n$, which give successively better lower bounds
on the private classical capacity.
{\bf (b)} $H(\cN(\proj{\ps_1}))$, the entropy of the output state
for a pure input or, equivalently, the entanglement between Eve's
output and Eve's environment. This is an intermediate quantity
required in the evaluation of $C_p^{(1)}$ and was also the focus
of~\cite{Alsing03}, where it was approximated by diagonalizing the
$k=0,1$ blocks of equation (\ref{Eve_state}).  Using our methods, it
can be shown that the exact value of the entanglement is $-3\left[{z
\ln z /(1-z)}+\ln{(1-z)}\right] +(1-z)^2\frac{\partial}{\partial z}
\frac{\partial}{\partial s} \Li{s}{z}|_{s=0}$ nats.}
 \label{fig:classical}
\end{figure}

Releasing the restriction, however, yields a non-zero capacity. In
fact, because $I(A;E_c)_\r$ in Theorem \ref{thm:qpriv} is a concave
function of $\ph^{A'}$ and the Unruh channel is covariant, the
maximum will be achieved with $\ph^{A'}$ maximally mixed. Evaluating
the formula yields a compact expression for $Q_p(\id,\cN)$ which we
have plotted in figure \ref{fig:classical}:
\begin{equation}
{1\over2}\left(1-{(1-z)^3\over2}{\partial^2\over\partial z^2}{\partial\over\partial s}
    \left[(z-1)\Li{s}{z}\right]_{s=0}\right),
\end{equation}
where $\Li{s}{z}$ is the polylogarithm function.

\section{Private classical capacity} \label{sec:cpriv}

Define the private classical capacity $C_p(\cN_1,\cN_2)$ as the
optimal rate, measured in bits per channel use, at which Alice can
send classical data to Bob over the channel $\cN_1$ in such a way
that Eve is incapable of distinguishing the messages based on her
view, the output of the channel $\cN_2$. This definition generalizes
the notion of private classical capacity introduced in
\cite{Devetak05}, which corresponds to the special case where
$\cN_2$ is the complement of $\cN_1$.

Using the methods of that paper and some additional arguments, one
can show that $C_p(\id_2,\cN) = \lim_{n\rar\infty}
C_p^{(n)}(\id_2,\cN)$, where $C_p^{(n)}(\id_2,\cN)$ is
\begin{equation}
 1 - H\big(\cN(I/2)\big) + \max_{\ket{\psi_n}}
   \frac{1}{n} H\big( \cN^{\ox n}(\proj{\psi_n}) \big)
\end{equation}
and $\ket{\psi_n}$ is any pure input state to $n$ copies of the
channel.
Evaluating the capacity therefore reduces to determining the maximal
output entropy of the Unruh channels $\cN^{\ox n}$ for pure input
states.
The optimization for $n=1$ is trivial due to the covariance of $\cN$
and gives that $C_p(\id_2,\cN)$ is bounded below by
\begin{multline}
C_p^{(1)}(\id_2,\cN) = (1-z)^2\frac{\partial}{\partial z}\left[\frac{\partial}{\partial s}\Li{s}{z}|_{s=0}\right]\\
-\frac{(1-z)^3}{2}\frac{\partial^2}{\partial
z^2}\left[z\frac{\partial}{\partial s}\Li{s}{z}|_{s=0}\right].
\end{multline}
We plot these bounds for $n=1,2$ in figure \ref{fig:classical}.

\section{Conclusions}

The assumption that an eavesdropper is accelerating can be exploited
to send data securely for all non-zero accelerations. In the case of
the private quantum data, we found a single-letter formula for the
capacity for general eavesdropper channels, demonstrating it to be
equal to the entanglement-assisted quantum capacity of the channel
to the eavesdropper's environment. We leave it as an open question
to explain why these seemingly unrelated tasks should have matching
capacity formulas but note that in light of \cite{Hastings08}, these
are now the only channel capacity problems in quantum information
that can be considered fully solved.\footnote{There is also a
remarkable formula for the so-called environment-assisted quantum
capacity of a quantum channel~\cite{SmolinVW05} but that problem is
of a very different type since it assumes full control of the
channel's environment, nearly the opposite of what is normally meant
by a noisy channel.}
In the case of
private classical data transmission, the problem of calculating the
associated private classical capacity reduces to that of determining
the maximal output entropy of the Unruh channel for pure input
states. This entropy corresponds to the entanglement between Rindler
field modes accessible to the eavesdropper and those not accessible,
a question of independent interest~\cite{Alsing03} resolved in this
paper.

When evaluating the private quantum capacity with an accelerating
eavesdropper, we began by considering isometric encodings, a class
known to be sufficient for non-private quantum data transmission.
With this restriction, private quantum data transmission reduces to
sending entanglement beyond the eavesdropper's horizon. An argument
related to the impossibility of cloning demonstrates this to be
impossible, an observation reminiscent of the analysis
in~\cite{Susskind93,Hayden07}, where the interplay of the no-cloning
theorem and horizons was used to place self-consistency constraints
on the black hole complementarity principle. We ended by evaluating
the private quantum capacity for unrestricted encodings, finding a
compact expression for the capacity which is non-zero for all
positive accelerations, in sharp contrast to no-go result for
isometric encoders.

\section*{Acknowledgments}
The authors thank John Preskill for
reminding us of the dangers of preliminary calculations and Jon Yard
for helpful conversations.
This work was
supported financially by the Canada Research Chairs program, CIFAR,
FQRNT, MITACS, NRO, NSERC, QuantumWorks and the Sloan Foundation.

\bibliographystyle{unsrt}
\bibliography{unruh_capacity}

\end{document}